# Origins of bad metal conductivity and the insulator-metal transition in the rare-earth nickelates

R. Jaramillo[1†*], Sieu D. Ha[1], D. M. Silevitch[2] and Shriram Ramanathan[1]

[1]School of Engineering and Applied Sciences, Harvard University, Cambridge, Massachusetts 02138, USA.

[2]The James Franck Institute and Department of Physics, The University of Chicago, Chicago, Illinois 60637, USA.

[*]Correspondence should be addressed to R. Jaramillo at rjaramil@mit.edu

**Abstract:** For most metals, increasing temperature (*T*) or disorder will quicken electron scattering. This hypothesis informs the Drude model of electronic conductivity. However, for so-called bad metals this predicts scattering times so short as to conflict with Heisenberg's uncertainty principle. Here we introduce the rare-earth nickelates ($R$NiO$_3$, $R$ = rare earth) as a class of bad metals. We study $SmNiO_3$ thin films using infrared spectroscopy while varying *T* and disorder. We show that the interaction between lattice distortions and Ni-O bond covalence explains both the bad metal conduction and the insulator-metal transition in the nickelates by shifting spectral weight over the large energy scale established by the Ni-O orbital interaction, thus enabling very low $\sigma$ while preserving the Drude model and without violating the uncertainty principle.

[†]Present address: Massachusetts Institute of Technology, Cambridge, Massachusetts 02139, USA.

For metals the electron-phonon scattering lifetime $\tau$ decreases with increasing $T$, with a linear dependence $\sigma^{-1} \propto \tau^{-1} \propto T$ at sufficiently high $T$. Elementary quantum theory dictates that this relationship cannot continue indefinitely. According to Heisenberg's uncertainty principle the uncertainty $\Delta E$ of a particle's energy is inversely proportional to its lifetime: $\Delta E = h/\tau$. Therefore there exists a minimum $\tau$ below which the concept of a well-defined quasiparticle energy becomes unphysical. This lower bound on $\tau$ implies a minimum metallic conductivity ($\sigma_{MIR}$). This bound is called the Mott-Iofe-Regel (MIR) limit and $\sigma_{MIR}$ is approximately equal to 1 $m\Omega^{-1}\cdot cm^{-1}$ [1,2]. Most metals reach their melting temperature well before the MIR limit. There are some so-called "saturating" metals for which $\sigma(T)$ approaches $\sigma_{MIR}$ and saturates, thus validating the MIR limit. However, in bad metals the relationship $\sigma^{-1} \propto T$ continues unabated through the MIR limit. According to the Drude model these metals have lifetimes so short that the quasiparticles should be unstable (*i.e.* decohere), producing an insulating state, and yet the transport properties remain metallic. Bad metal conductivity is often found in strongly correlated materials such as the high $T$ superconductors and the Mott insulator-metal transition system $VO_2$. The phenomenon of bad metal conductivity is a central problem in condensed matter physics [1–3].

Here we study bad metal conductivity and the insulator-metal transition in the rare-earth nickelates ($R$NiO$_3$, $R$ = rare earth). The nickelate phase diagram features an antiferromagnetic insulator at low $T$ and a paramagnetic metal (PM) at high $T$. For $R$ = Sm and heavier there is also an intermediate paramagnetic insulator for $T_N < T < T_{IM}$ ($T_N$ = Néel temperature, $T_{IM}$ = IMT temperature)[4]. $T_{IM}$ and the measure of correlation in the metal phase (*e.g.* the suppression of the coherent spectral weight) can be tuned by cation substitution, epitaxial strain, or dimensionality control[4–6]. The nickelates have strong interactions between the charge, spin, orbital and lattice degrees of freedom, and this complexity has obscured the physics of the IMT. Several distinct scenarios have been proposed including a Mott transition, charge and spin density wave formation, orbital order, and a polaronic insulator[7–13]. Experiments show that the metallic phase is consistent with strong electronic correlations and that the electron-phonon interaction is important, but have not presented a clear explanation of the IMT[5,7,10,13]. Recently, both total-energy theory and a model Hamiltonian theory have partially reproduced the phase diagram[14,15].

To directly address the apparent breakdown of the lifetime hypothesis in a bad metal we measure a series of samples with controlled levels of disorder. We grow $SmNiO_{3-\delta}$ epitaxial thin films on $LaAlO_3$ by high pressure sputtering. By varying the sputtering pressure we can control the concentration of oxygen vacancies, producing films with different amounts of disorder[16,17]. Here we report results for five samples, here named SNO1-SNO5, with increasing amounts of disorder. SNO1 is the closest to stoichiometric and displays a sharp IMT. SNO5 is the farthest from stoichiometric and displays a broadened IMT.

In Fig. 1a we present $\sigma_{DC}(T)$ for a conventional metal (Cu), a saturating metal ($Ti_{0.89}Al_{0.11}$), and $R$NiO$_3$. Cu is a good conductor at all $T$: both $\sigma_{DC}$ and the fractional slope -dln($\sigma$)/d$T$ (Fig. 1b) are large[18]. $Ti_{0.89}Al_{0.11}$ is a saturating metal: $\sigma_{DC}$ saturates somewhat above $\sigma_{MIR}$ at high $T$, and the slope -dln($\sigma$)/d$T$ is 30 times smaller than for Cu [1]. We also plot our measurements on SNO3 at high $T$ along with previously published data for $R$NiO$_3$, all in the PM phase. The published data include $R$ = La, Pr, Nd, Sm, Gd, and doped compounds. For all $R$NiO$_3$ $\sigma_{DC}$ is below $\sigma_{MIR}$ at high $T$. Importantly, the fractional slope -dln($\sigma$)/d$T$ is large for all $R$NiO$_3$. For SNO3 the fractional slope at 800 K is nearly identical to that of Cu, although $\sigma_{DC}$ is orders of magnitude lower. This means that the $\sigma_{DC}$ does not saturate even as it falls below the MIR limit. Fig. 1 establishes $R$NiO$_3$ as a class of bad metals.

We probe the complex IR conductivity ($\sigma(E) = \sigma'(E) + i\sigma''(E)$) in the range $E$ = 25-1550 meV for $T$ between 295 and 415 K. We measured and analyzed 69 complete spectra for varying $T$ and disorder. We show in Fig. 2 the $T$-dependence of $\sigma'(E)$ for SNO1. The IMT is accompanied by a large increase in spectral weight below 400 meV, only some of which is transferred from the mid-IR peak near 550 meV. The Ni-O bending and stretching modes (near 35 and 75 meV, respectively) are narrow phonons for $T < T_{IM}$ but merge into the coherent quasiparticle response in the PM phase, becoming strongly damped for $T > T_{IM}$. We identify these phonons by reference to a general analysis of IR-active phonons in distorted (*Pnma*) perovskites[19]. The Ni-O stretching mode is associated with a breathing distortion of the NiO$_6$ octahedra, and the bending mode affects the Ni-O-Ni bond angle[11]. The quasiparticles in the PM phase are polarons[10,11]; our data identify the phonon modes that strongly couple with the electronic degrees of freedom. The emergence of a strongly damped phonon response for $T > T_{IM}$ also explains the far-IR minimum in $\sigma'(\omega)$, which is present in our data and is commonly observed in manganites and bad metals[2].

The mid-IR peak position (Fig. 2, inset) has the $T$-dependence expected for an order parameter and has been identified with $E_g$ [20]. However, instead of disappearing in the PM phase it persists as a shoulder on the broad quasiparticle response. We identify this shoulder as a Holstein side band, which is associated with the breakup of a polaron into its constituent phonon and electron degrees of freedom[21]. The connection of the $E_g$-like feature for $T < T_{IM}$ to the Holstein side band for $T > T_{IM}$ suggests that the IMT is driven by the freezing of the same lattice distortion that is responsible for polaron formation in the PM phase. This is consistent with experiments showing a reduction in symmetry from orthorhombic for $T > T_{IM}$ to monoclinic for $T < T_{IM}$ and with theory showing that the insulating phase is controlled by a period doubling lattice distortion[14,15,22,23].

The effects of varying $T$ and disorder are captured by the spectral weight:

$$f(E) = \frac{4}{\varepsilon_0 h}\int_0^E dE'\sigma'(E'). \tag{1}$$

$f(E)$ counts the quasiparticles that contribute to conductivity at frequencies $\omega' < E/\hbar$. $f(E \approx W)$, where $W$ is the conduction band width, measures the effect of strong coupling on the electronic structure. The plasma frequency ($\Omega_p$) is:

$$\Omega_p = \sqrt{\frac{nq^2}{m^* \varepsilon_0}} = \sqrt{f(W)}\,. \qquad (2)$$

Strong coupling is taken to be any interaction that significantly alters $n$ or $m^*$ in the PM phase.

We plot in Fig. 3a the $T$-dependence of $f(E)$ for SNO1. The IMT is accompanied by a steep rise in $f(E)$ in the range $E < 400$ meV, signaling the collapse of $E_g$ and the filling of the conduction band. Much of the spectral weight transfer comes from high energy states beyond 1.5 eV, as observed for other $R$NiO$_3$ [5,7]. For $T > T_{IM}$ $f(E)$ is nearly identical to that measured on 2.9% tensile strained $LaNiO_3$-on-$DyScO_3$ at room $T$ [5]. This suggests a common mechanism for spectral weight suppression for epitaxial strain and rare earth substitution in $R$NiO$_3$.

We plot in Fig. 3b-c results from an extended Drude analysis of all samples (see Supplementary Discussion). For an IMT driven by electron-electron correlation $m^*$ diverges as $T \rightarrow T_{IM}^+$, as observed in Mott insulators such as $VO_2$ [3]. For $SmNiO_3$ we observe no such mass enhancement approaching $T_{IM}$. In a conventional metal the effect of adding disorder is to decrease $\tau$, thereby lowering $\sigma_{DC}$. For $SmNiO_3$ we see that $\tau(\omega \rightarrow 0)$ is nearly constant for all samples, irrespective of the amount of disorder.

According to the Drude model if both $m^*$ and $\tau$ are constant then a changing $\sigma_{DC}$ must result from a changing $n$. In Fig. 4a we overlay $\sigma_{DC}(T)$ for SNO1-5 with $\Omega_p^2$ (see Supplementary Discussion). $\Omega_p^2$ responds to changes in $T$ and disorder in nearly the same way as $\sigma_{DC}(T)$. The correspondence in the PM phase is remarkable: $\Omega_p^2$ decreases in response to increased $T$ or disorder, just as $\tau$ does in a conventional metal (Fig. 4b). Fig 4 suggests an explanation for bad metal conductivity in the nickelates: small changes in $T$ or disorder produce continuous shifts of spectral weight over a large energy scale. This stands in stark contrast to normal metals in which changes in $T$ or disorder affect $\tau$. The defining characteristic of bad metals may be the presence of a strong interaction that can continuously shift spectral weight between the conduction band and high energy states in response to small changes in $T$ and disorder, yielding metallic conductivity below $\sigma_{MIR}$ while preserving the Drude model and without violating the uncertainty principle [2].

The interaction between the Ni-O covalence and the electron-phonon interaction is most directly responsible for both the bad metal conductivity and the IMT. The ground state electronic configuration of $R$NiO$_3$ is nominally $t_{2g}^6 e_g^1$ ($Ni^{3+}$). However, this ionic representation is inaccurate due the strong covalent sigma bonds between Ni $e_g$ and O 2p orbitals, and $R$NiO$_3$ are best described as negative charge-transfer valence bond insulators[24,25]. Theory shows that the low-energy density of states (DOS) near $E_F$ is antibonding in character, the bonding DOS is centered 6-8 eV below $E_F$, and this separation is the dominant high energy feature in the electronic structure[14,26]. Importantly, Ni-O covalence is controlled by lattice distortions (Fig. 4c-d)[24]. Overlap between the Ni $e_g$ and O 2p orbitals depends on the Ni-O bond lengths and the Ni-O-Ni bond angles. The bond lengths are affected by the breathing mode, and the bond angles are affected by the bending mode. The high energy electronic structure of $R$NiO$_3$ is therefore controlled by the local geometry of the NiO$_6$ octahedra.

Both the bending and the breathing modes condense at $T_{IM}$, resulting in a discrete change in the Ni-O-Ni bond angle and a disproportionation into expanded (Ni1 sites) and contracted (Ni2 sites) octahedra[13–15,22,23]. This affects the Ni-O covalence and transfers spectral weight over large energies, as observed experimentally[7]. In the insulating phases the Ni1 sites are closer to the ionic limit $t_{2g}^6e_g^2\underline{L}$ ($\underline{L}$ denotes a "ligand hole" on an oxygen orbital), with spin quantum number $S = 1$. The Ni2 sites are closer to the covalent limit, with electrons in bonding orbitals forming $S = 0$ singlets[12–14]. $E_g$ is approximately 10 times smaller than the large energy scales in the system, and results from period doubling that splits the low energy DOS[8,12,14]. The low energy DOS is predominantly $\underline{L}$-like; transport through the $e_g^2$ manifold is suppressed by the electron-electron Coulomb energy. In the PM phase spectral weight is shifted by dynamic electron-phonon interactions, as suggested by the strongly damped bending and breathing modes and by the observation of dynamic Ni1-Ni2 fluctuations above $T_{IM}$ [23]. It is a challenge to theory to further quantify the dynamic electron-phonon interaction in the PM phase. The full description will likely include a pseudogap in the metallic DOS.

Although the electron-phonon interaction is paramount, $R$NiO$_3$ are not polaronic insulators in the conventional sense. In a polaronic insulator coherent charge transport breaks down at high $T$ because electron-phonon scattering suppresses the bandwidth to the point where the quasiparticle concept fails. Polaronic insulators such as the colossal magnetoresistive manganites are characterized by a high $T$ insulating phase with hopping conductivity and often short range orbital order[28]. In contrast the insulating phase of $R$NiO$_3$ is found at low $T$, does not exhibit clear hopping conductivity (see Supplementary Discussion), and is characterized by a coherent lattice distortion without orbital order[29]. In the language of Goodenough, $R$NiO$_3$ are "collective electron insulators" in which the electron-phonon interaction is strong enough to suppress the bandwidth, allowing the DOS to be split by period doubling order for $T < T_{IM}$, but not strong enough to destroy the coherent quasiparticles[30,31].

In this paper we identify $R$NiO$_3$ as a class of bad metals, and we use IR spectroscopy to explain both the bad metal conductivity and the IMT in SmNiO$_3$ thin films in terms of the interaction between lattice distortions and the Ni-O bond covalence. In the PM phase spectral weight is continuously redistributed in response to changes in $T$ or disorder. This is consistent with explanations of bad metal conductivity in Mott-Hubbard materials such as the high $T_c$ copper oxides[2]. We conclude that a strong interaction that can shift spectral weight over a large energy scale in response to small changes in $T$ or disorder is essential, but the identity of this interaction may vary. It would be interesting to study the effects of controllably introducing different types of disorder on the spectral weight and the lattice distortion in the PM phase.

**Acknowledgments**

The authors acknowledge ARO MURI (W911-NF-09-1-0398) and the NSF (DMR-0952794 and DMR-1206519) for financial support. This work was performed in part at the Center for Nanoscale Systems which is supported by the NSF under award no. ECS-0335765, at shared facilities of the University of Chicago MRSEC which is supported by the NSF under award DMR-0820054, and at the MRSEC Shared Experimental Facilities at MIT which is supported by the NSF under award number DMR-08-19762. R.J. particularly acknowledges technical assistance from Timothy McClure at the MIT MRSEC Shared Experimental Facilities.

**Author Contributions**

RJ conceived of, planned, and executed the experiments, analyzed and interpreted the results, and wrote the manuscript. SDH grew the samples and edited the manuscript. DS performed the low temperature electronic transport measurements. SR provided support and technical discussions.

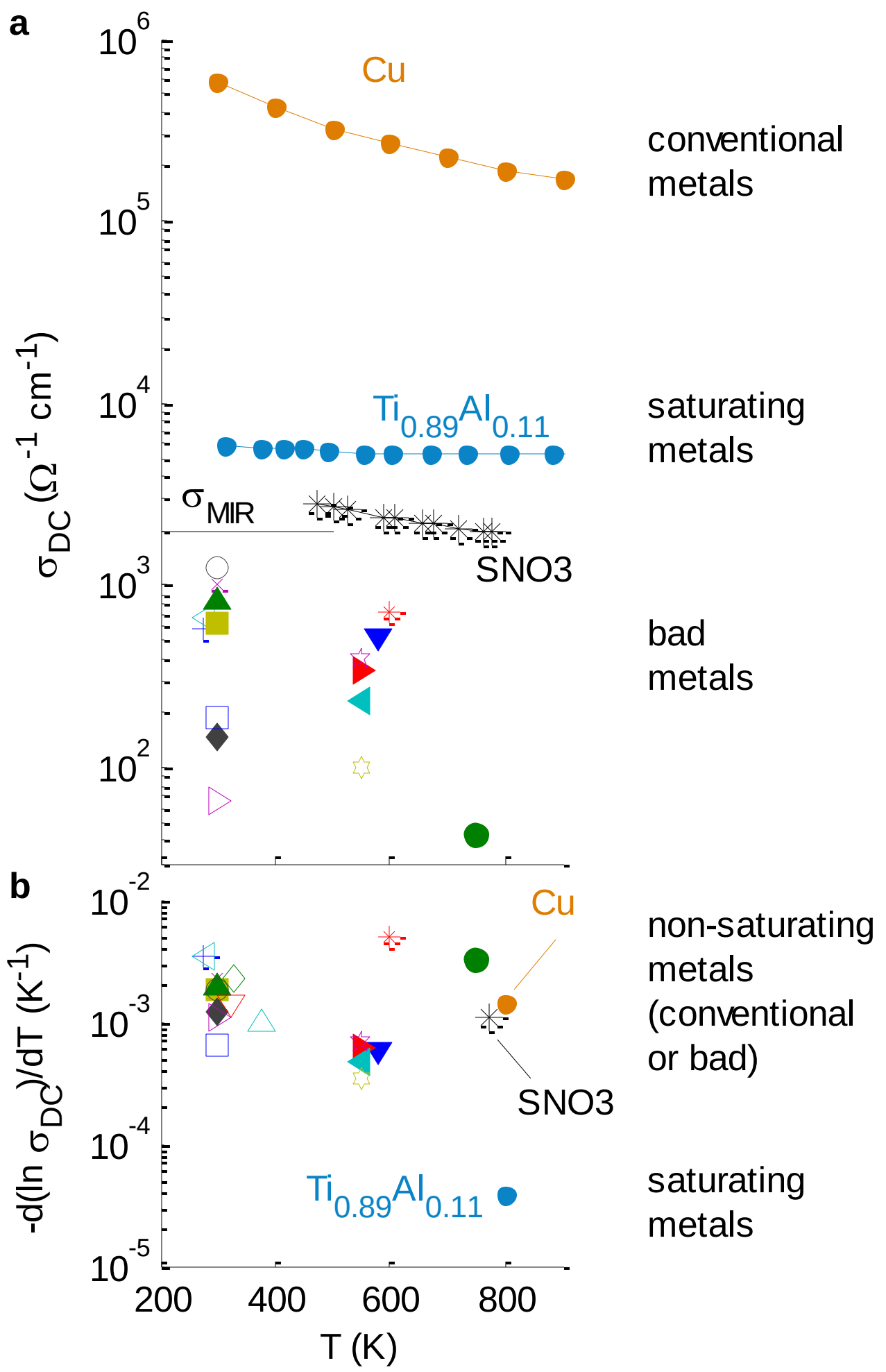

**Figure 1:** Bad metal conductivity in the high *T* PM phase of the nickelates. (✳): our data for SNO3. (+,○,□): $LaNiO_3$. (☆,☆): $La_{1-x}Pb_xNiO_3$. (×): $PrNiO_3$. (◁,■,◆,▲,▷): $NdNiO_3$. (◇,▽,△): $Sm_xNd_{1-x}NiO_3$. (▶,◀): $SmNi_{1-x}Co_xO_3$. (✳,▼): $SmNiO_3$. (●): $GdNiO_3$. We also show data for a conventional metal (Cu, -●-)[18], and a saturating metal ($Ti_{0.89}Al_{0.11}$, -●-)[1]. References for published data on $R$NiO$_3$ are listed in the Supplementary Discussion, and the points represent the highest reported *T*. (a) $\sigma_{DC}$ at high *T*. There is a clear separation of scales between good metals such as Cu, saturating metals such as $Ti_{0.89}Al_{0.11}$, and bad metals such as $R$NiO$_3$. For $R$NiO$_3$ we estimate $\sigma_{MIR} \approx 2000\ \Omega^{-1}\ cm^{-1}$ (see Supplementary Discussion). (b) Fractional derivative $-d\ln(\sigma)/dT$ corresponding to the data in (a). Both the bad metals $R$NiO$_3$ and the conventional metal Cu have a relatively large derivative at high *T*. In contrast, the saturation of $\sigma_{DC}$ for $Ti_{0.89}Al_{0.11}$ results in a fractional derivative that is suppressed by a factor of 10-100 compared to Cu and $R$NiO$_3$.

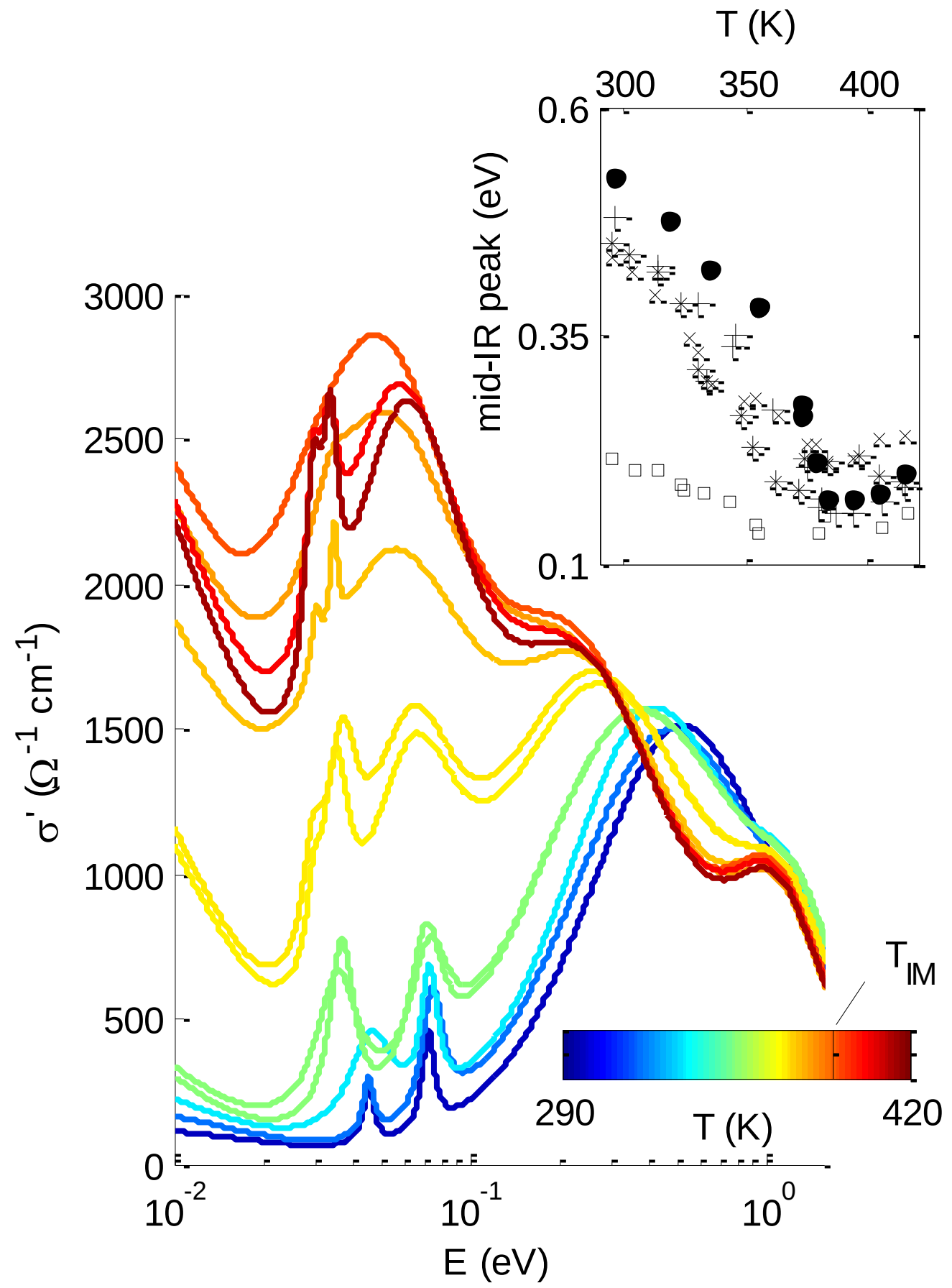

**Figure 2:** $\sigma'(E)$ measured across the IMT for SNO1. $T_{IM}$ as measured by the change in slope of $\sigma_{DC}(T)$ is marked on the color bar. (Inset) $T$-dependence of the mid-IR peak position for all 5 samples SNO1, SNO2,…SNO5 (●, +, ✳, ×, □).

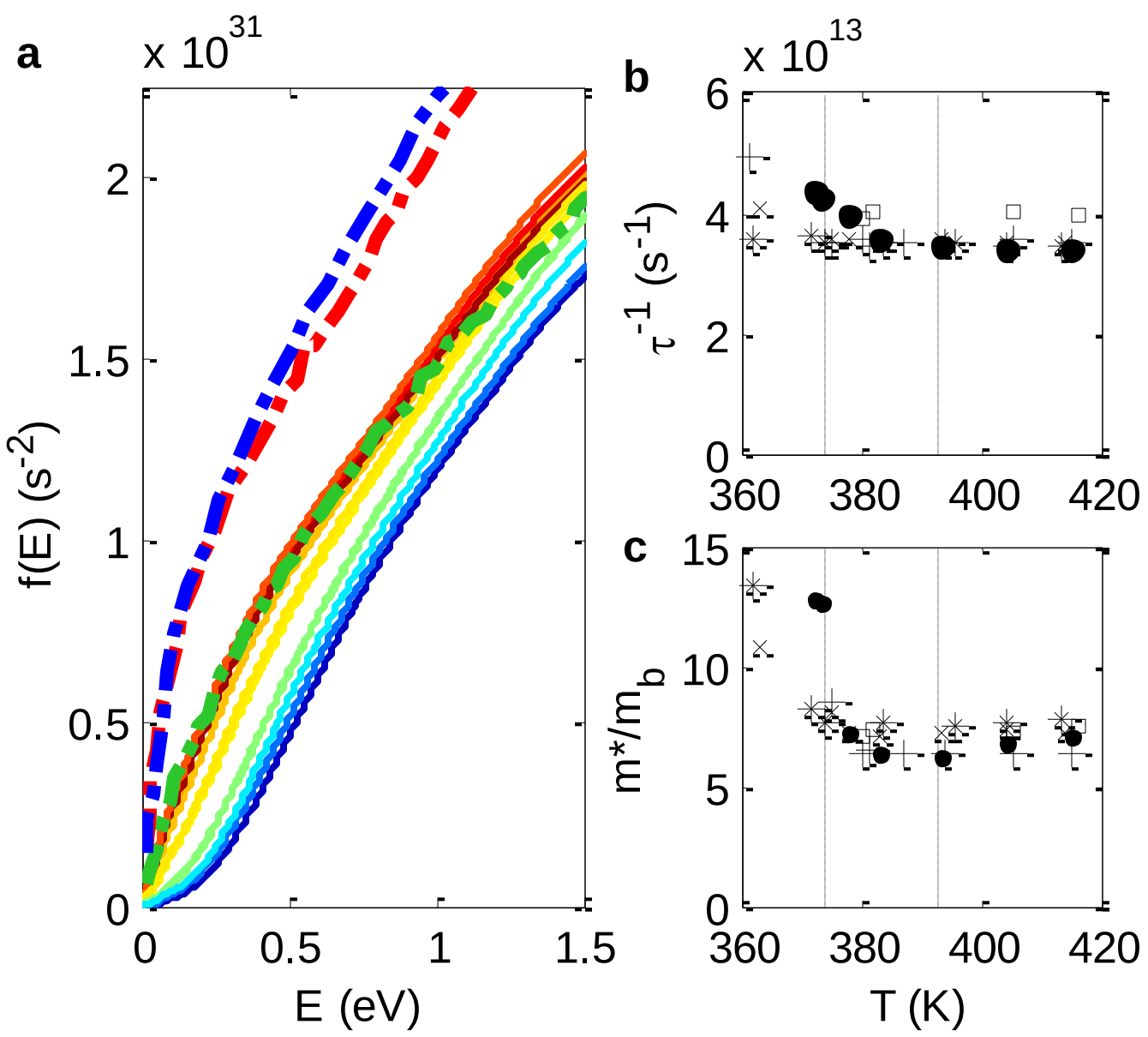

**Figure 3:** Spectral weight and extended Drude analysis. (a) Solid lines: $f(E)$ for SNO1 across the IMT; color scale is the same as Fig. 2. Broken lines: $f(E)$ for epitaxially strained $LaNiO_3$ [5]; red = -1.1% strain on $LaAlO_3$, blue = +0.6% strain on $(LaAlO_3)_{0.3}(Sr_2AlTaO_6)_{0.7}$; green = +2.9% strain on $DyScO_3$. (b,c) Extended Drude analysis for SNO1-SNO5. Symbols are the same as in Fig. 2. Data is significant only in the PM phase ($T > T_{IM}$). $T_{IM}$ = 394, 386, 380, 387, and 374 K for SNO1, SNO2,…SNO5, respectively. This range of $T_{IM}$ values is shown by the dashed vertical lines. (b) $\tau^{-1}(\omega \rightarrow 0)$ is nearly constant for all samples, irrespective of $T$ or disorder. (c) $m^*(\omega \rightarrow 0)/m_b$ is $T$-independent for $T > T_{IM}$ for all samples (note the variation in $T_{IM}$).

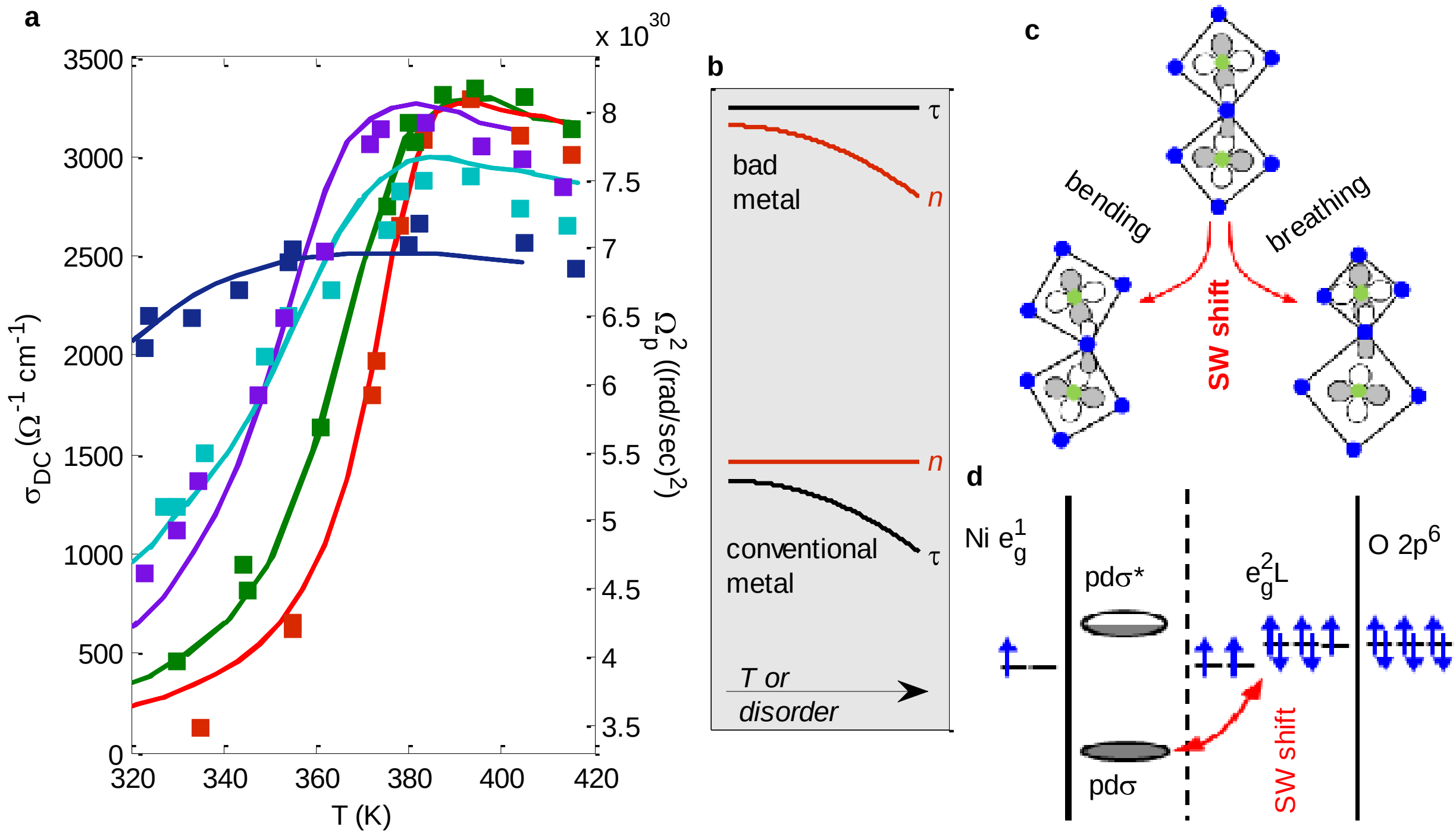

**Figure 4:** Spectral weight shifts in response to *T*, disorder, and the IMT. (a) $\sigma_{DC}$ (left axis, lines) and $\Omega_p^2$ (right axis, squares) for SNO1, SNO2,… SNO5 (red, green, purple, teal, blue) as a function of *T* across $T_{IM}$. In the PM phase $\Omega_p^2$ decreases in response to increased *T* or disorder, just as τ does in a conventional metal. (b) Schematic showing the different responses of a bad metal and a conventional metal to increases in *T* or disorder. (c) Schematic of the bending and breathing distortions that shift spectral weight by altering the Ni-O covalence. Blue = O, green = Ni. We draw select Ni $e_g$ and O 2p orbitals to illustrate how the distortions affect orbital overlap. (d) Schematic of the electronic structure; emphasis is on Ni-O bonding and how the Ni $e_g$ and O 2p orbitals bond in the covalent and ionic limits. Outer columns correspond to the electronic structure in the nominal $t_{2g}^6 e_g^1$ configuration. The inner-left column shows the limit of covalent Ni-O bonding, with a high energy separating the pdσ and pdσ* states[14,26]. The inner-right column shows the limits of ionic Ni-O bonding, with the $e_g^2$ and $\underline{L}$ states separated by the (negative) charge transfer energy.

# Supplementary Discussion

## 1. Analysis of spectral weight and the extended Drude model

In order to estimate the plasma frequency ($\Omega_p$) of our SNO films we need to choose a cutoff energy (*W*) for the spectral weight integral, Eq. 1. This should correspond to the electronic bandwidth in the PM phase, so that coherent excitations are included but incoherent transitions are excluded. Coherent excitations are transitions between relatively long-lived quasiparticle states that contribute directly to the transport and thermodynamic properties of the metal phase. Incoherent excitations are higher energy transition between states with short lifetimes, such as between the valence and conduction band or between the lower and upper Hubbard band. Different values for *W* can be found in the literature on $RNiO_3$. For example, Katsufuji uses $W$ = 0.3 eV, determined by the intersection of $\sigma'(\omega)$ for 290 K and 9 K (the "isobetic" point) for $NdNiO_3$ [20]. Ouellette uses $W$ = 0.2 eV, determined by the qualitative turnover in the $f(\hbar\omega)$ data for $LaNiO_3$ [5]. For our data we find that these two criteria lead to the same value $W \approx 3000$ cm$^{-1}$ (or 370 meV, as reported in the main text). This is illustrated in Fig. S1. Fig. S1a shows $\sigma'(\omega)$ for SNO1, highlighting the crossover between curves near 3000 cm$^{-1}$. Fig. S1b shows the derivative $df/d\lambda^{-1}$ for SNO1 at 120 °C. We take *W* to be the point halfway down the broad decrease in $df/d\lambda^{-1}$; this quantifies the qualitative criterion used by Ouellette in Ref. [5]. Both of these estimates lead to the choice $W$ = 3000 cm$^{-1}$ (372 meV).

We emphasize that our results reported here, including the extended Drude analysis in Fig. 3 and the plasma frequencies plotted in Fig. 4, are not qualitatively affected by the choice of *W*. Our conclusions are unchanged for choices of *W* down to at least 1500 cm$^{-1}$ (186 meV).

Sometimes in the FTIR literature the phonon contributions to $\sigma'(E)$ are excluded from the integral $f(E)$. This is inappropriate for a polaronic conductor such as $SmNiO_3$. The bending and breathing phonon modes are broad and strong for $T > T_{IM}$, and we argue they have clearly merged with the coherent electronic response. Therefore it is appropriate to include their contribution to $f(E)$.

The importance of including the contribution of the broad phonon modes to $\Omega_p$ is made clear in Fig. S2 where we consider the effect of neglecting the Drude term entirely. In Fig. S2a we reproduce Fig. 4, showing the correspondence between $\Omega_p$ and $\sigma_{DC}$ in the PM phase for varying *T* and disorder. In Fig. S2b we show a similar plot, but for each point $\Omega_p$ is calculated without the contribution from the Drude term. Even without the Drude contribution, $\Omega_p$ and $\sigma_{DC}$ correspond in the PM phase for varying *T* and disorder. This clearly demonstrates that the broad phonon modes contribute to the electrical conductivity in the PM phase, and therefore should be included in the coherent electronic response.

The extended Drude model is calculated from the complex conductivity $\sigma(E)$ and $\Omega_p$ [3]. It yields a frequency-dependent lifetime $\tau(\omega)$ and mass renormalization $m^*(\omega)/m_b$. $\tau(\omega\rightarrow 0)$ and $m^*(\omega\rightarrow 0)$ are the lifetime and the fully renormalized effective mass that appear in the Drude expression for $\sigma_{DC}$. $m_b$ is the so-called band mass that describes the dispersion of the quasiparticle bands in the absence of frequency-dependent scattering processes. In practice this often means the effective mass in the absence of strong correlations.

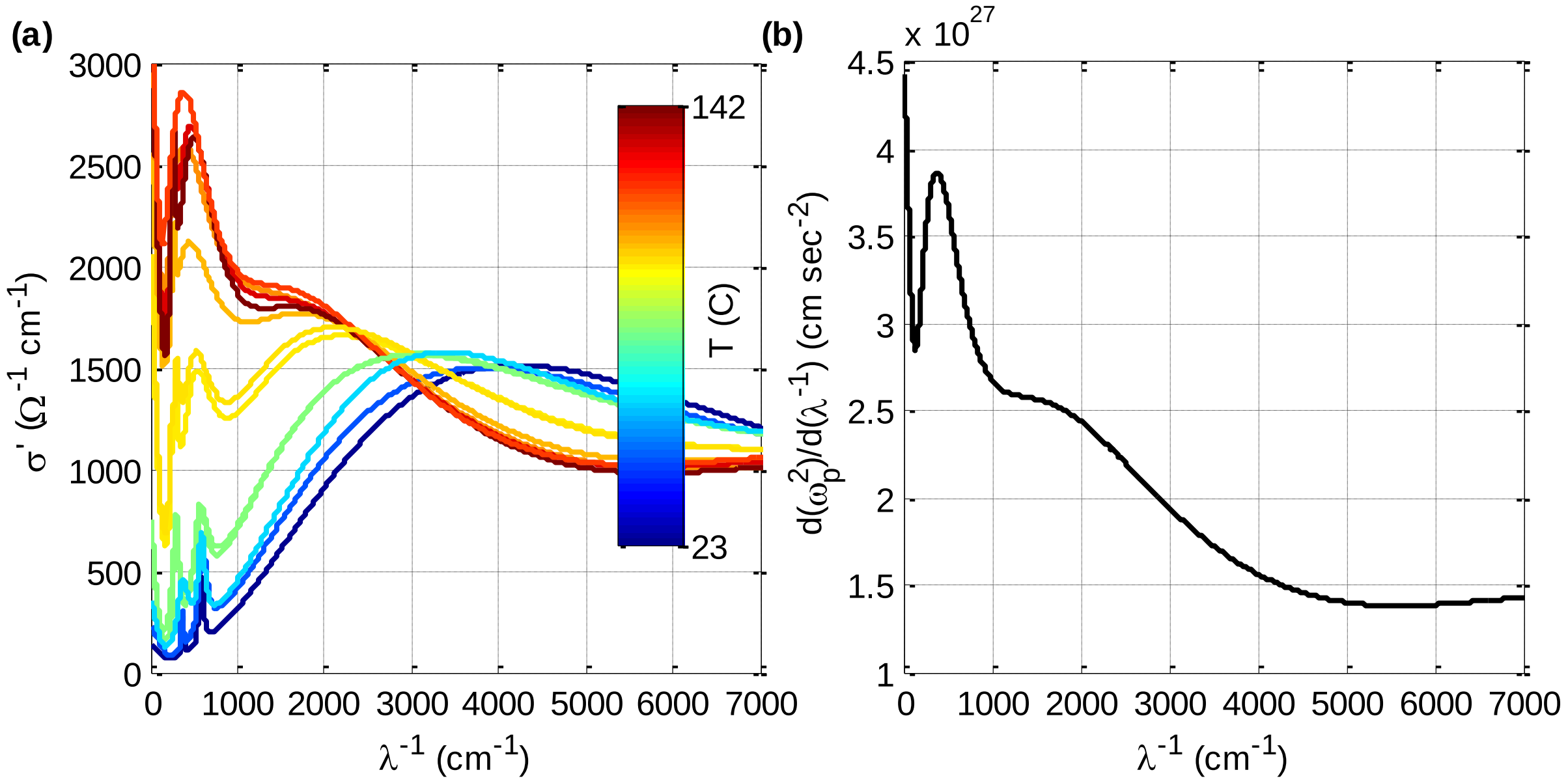

**Figure S1:** Choosing the cutoff ($W$) for the spectral weight integral. (a) $\sigma'(\omega)$ for SNO1. The low $T$ and high $T$ data intersect near 3000 $cm^{-1}$. This is called the isobetic point and is often taken as the bandwidth $W$ in the PM phase. (b) Derivative of the integrated SW $f(\lambda^{-1})$ for SNO1 at 120 °C ($f(\hbar\omega)$ is plotted in Fig. 3). The broad rise below 5000 $cm^{-1}$ is taken to be the onset of coherent excitations, and the halfway point of this rise at 3000 $cm^{-1}$ is an estimate of $W$.

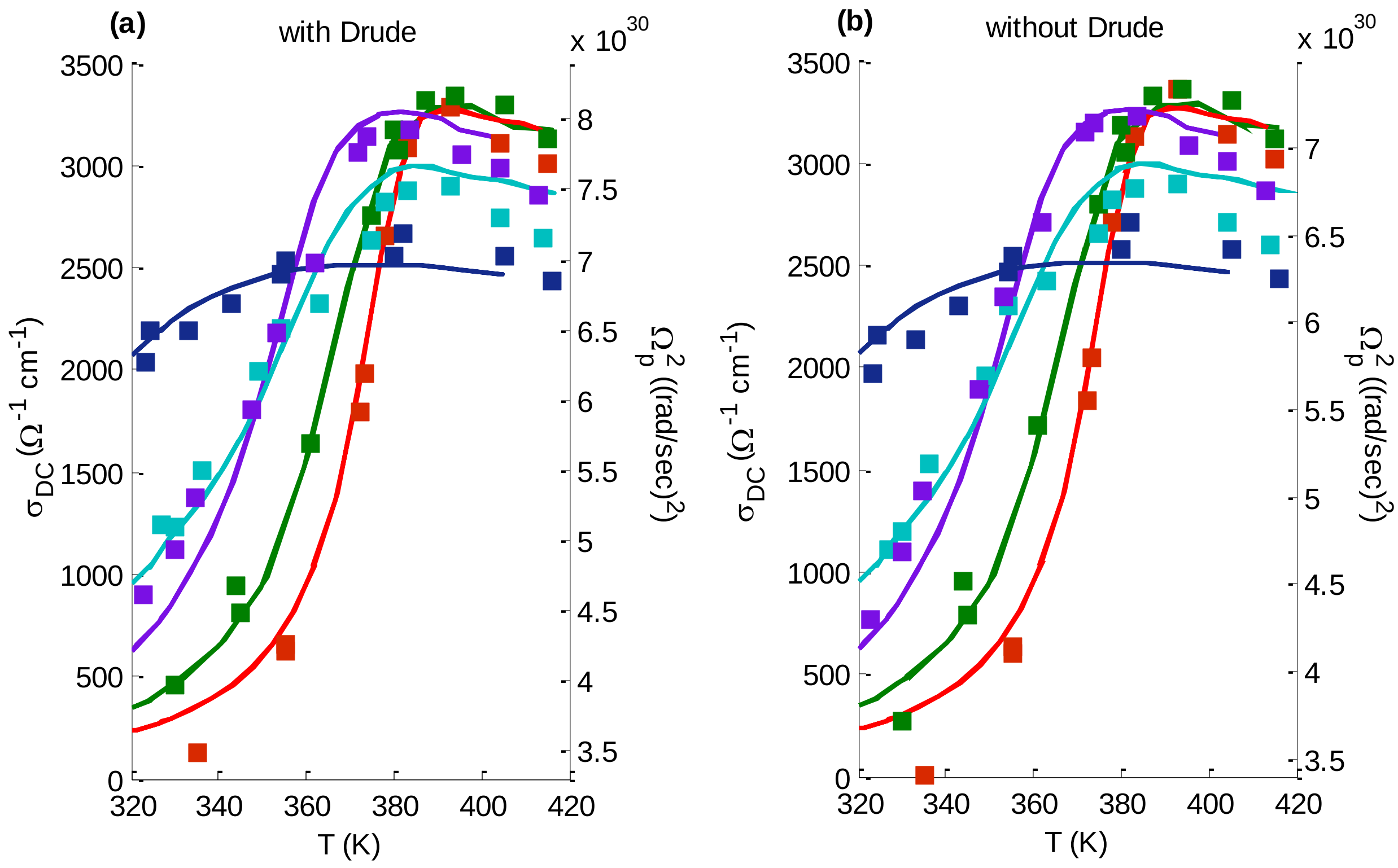

**Figure S2:** The importance of including the contribution of the broad phonon modes to $\Omega_p$ (a) Same as Fig. 4, showing the correspondence between $\Omega_p$ and $\sigma_{DC}$ in the PM phase for varying $T$ and disorder. $\sigma_{DC}$ are plotted as solid lines, $\Omega_p$ are plotted as squares. Colors correspond to SNO1, SNO2,… SNO5 (red, green, purple, teal, blue). (b) Similar to (a), but for each point $\Omega_p$ is calculated without the contribution from the Drude term. Even without the Drude contribution, $\Omega_p$ and $\sigma_{DC}$ correspond in the PM phase for varying T and disorder.

## 2. Correspondence between $\sigma(\omega\rightarrow 0)$ and $\sigma_{DC}$, and effect of including $\sigma_{DC}$ in the FTIR fits

We model our FTIR data for $E \geq 25$ meV (or $\lambda^{-1} \geq 200$ cm$^{-1}$). Because of this cutoff and in order to produce the best available dielectric models for our SNO films we have included the independently measured $\sigma_{DC}(T)$ in the modeled datasets. However, it is important to study how this finite energy cutoff and the inclusion of $\sigma_{DC}$ in the fitting procedure affect our results. In Figs. S3-S5 we present the results of fitting the data without including the measured $\sigma_{DC}(T)$ in the modeled datasets. We see some quantitative changes, but the main results of this work are not at all affected.

In Fig. S3 we plot $\sigma'(E)$ and $f(E)$ for SNO1 calculated by fitting the FTIR data with and without including the measured $\sigma_{DC}(T)$ in the modeled datasets. The differences in $\sigma'(E)$ are very subtle within the plotted energy window ($E$ > 10 meV), except for the highest temperature dataset. The differences in $f(E)$ are almost imperceptible.

In Fig. S4 we compare the measured $\sigma_{DC}(T)$ to the zero-energy extrapolation $\sigma'(\omega \rightarrow 0)$ of FTIR fits with and without including the measured $\sigma_{DC}(T)$ in the modeled datasets for SNO1 and SNO5. For SNO1 the correspondence is very good except at the highest $T$. The failure of the fit without $\sigma_{DC}(T)$ to accurately extrapolate to the measured $\sigma_{DC}(T)$ at high $T$ does not affect our results or conclusions. For SNO5 the correspondence is good at all $T$.

In Fig. S5 we compare $\Omega_p^2$ and the extended Drude analysis with and without including the measured $\sigma_{DC}(T)$ in the modeled datasets. For $\Omega_p^2$ the choice of whether to include $\sigma_{DC}(T)$ makes almost no difference. For the extended Drude analysis we do see small quantitative differences, but no qualitative changes to our results. The scattering rate $\tau^{-1}$ in the PM phase remains approximately equal for both SNO1 and SNO5. The mass enhancement $m^*/m_b$ remains flat near and above $T_{IM}$. The highest $T$ datapoint for SNO1 does show a rather large deviation in $m^*/m_b$: it changes from 6.5 at 131 °C to 2.3 at 142 °C. This change is too large and too abrupt to be physically meaningful, and reflects the difficulty of modeling the full AC conductivity ($\sigma(\omega) = \sigma'(\omega) + i\sigma''(\omega)$) with data that only extend to 25 meV.

In brief, the main results of this work are unaffected by the choice to include the independently measured $\sigma_{DC}(T)$ in the modeled datasets.

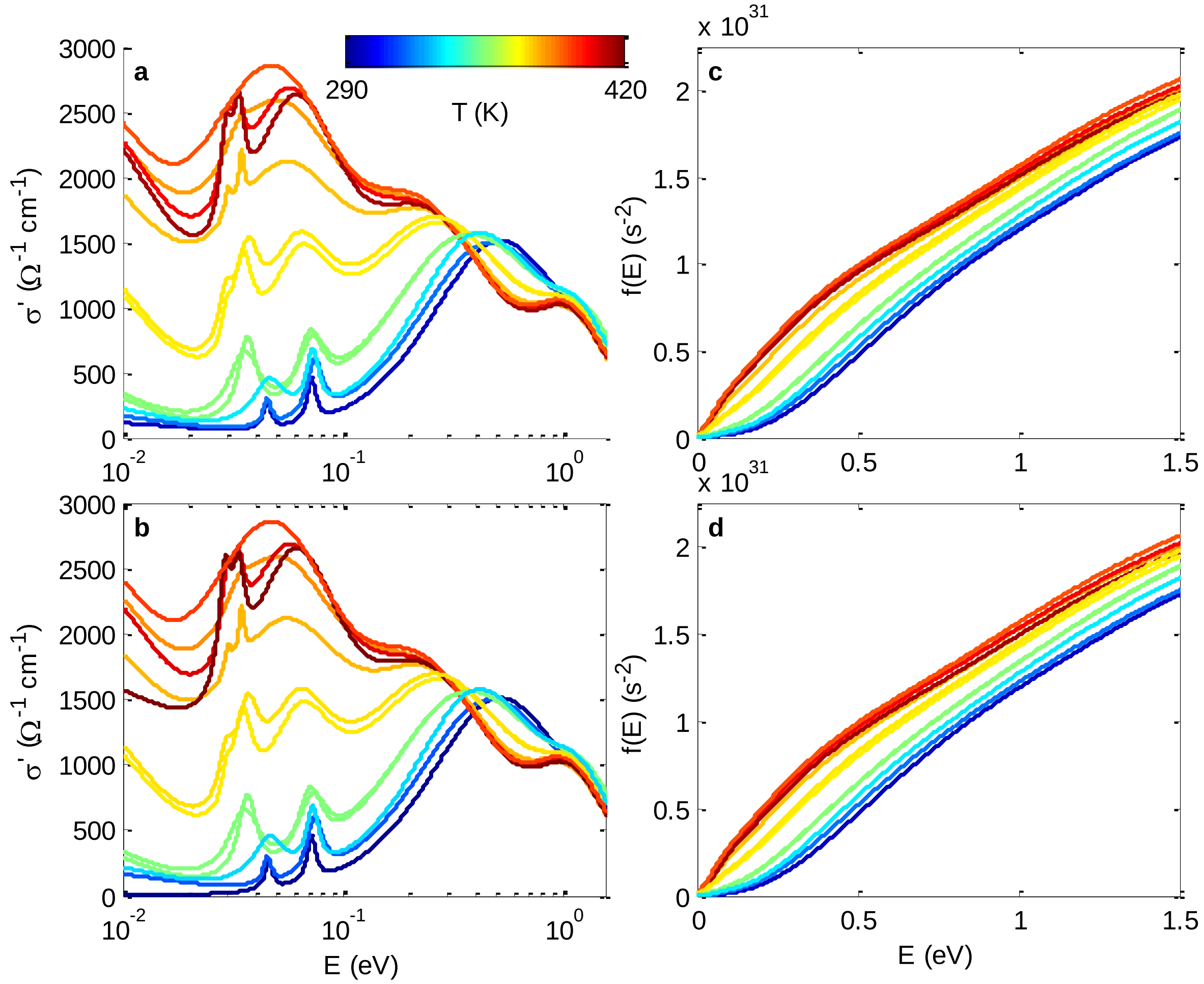

**Fig. S3:** $\sigma'(E)$ and $f(E)$ for SNO1 from FTIR fits with and without including the measured $\sigma_{DC}$ in the modeled datasets. (a-b) $\sigma'(E)$. (c-d) $f(E)$. Top row: fits including $\sigma_{DC}$ in the modeled datasets (same as in Fig. 2-3). Bottom row: fits not including $\sigma_{DC}$.

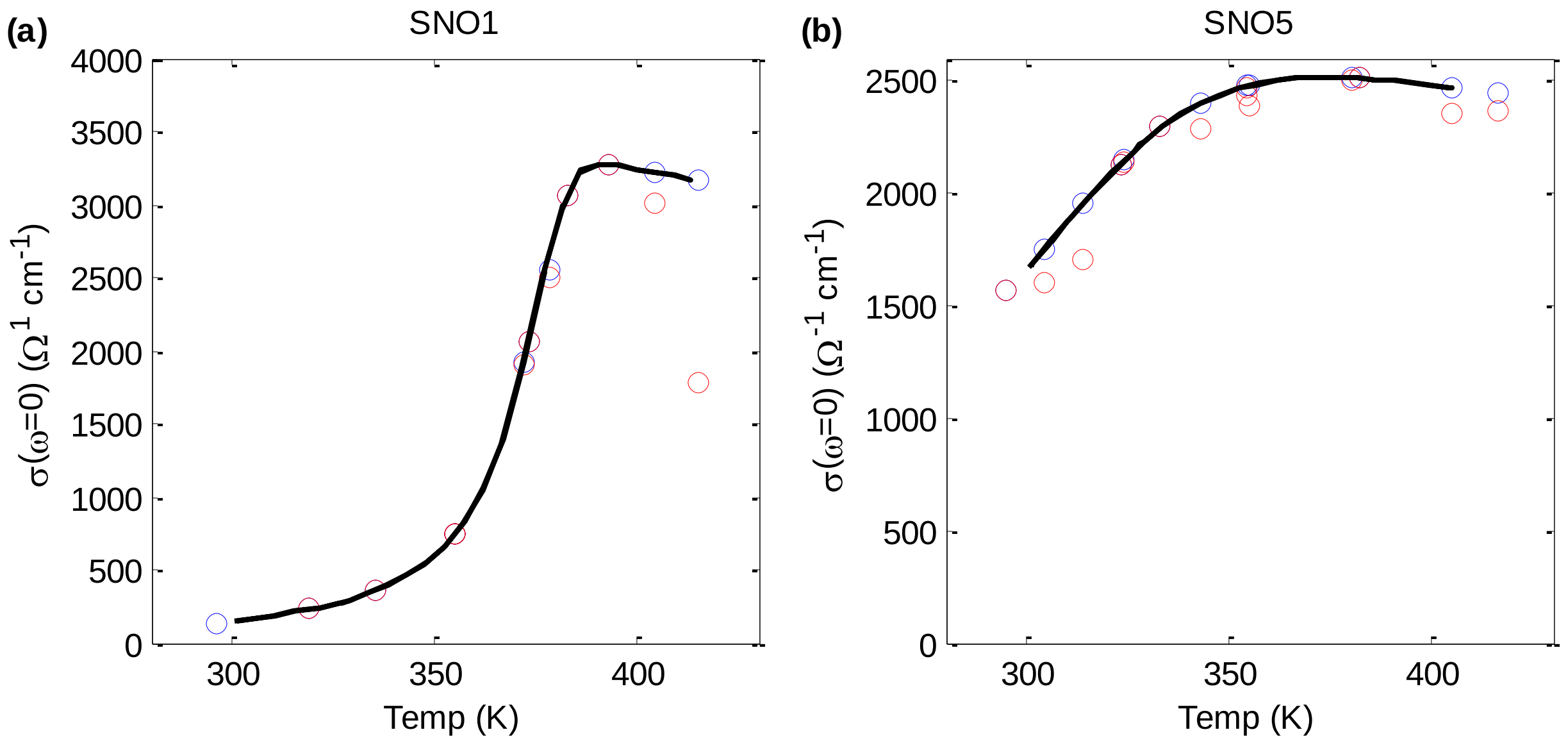

**Figure S4:** Correspondence between directly measured $\sigma_{DC}$ and the extrapolated $\sigma(\omega \rightarrow 0)$ from the FTIR fits for (a) SNO1 and (b) SNO5. Black curves show directly measured $\sigma_{DC}$. Blue circles show results for FTIR fits that include the measured $\sigma_{DC}$ in the modeled datasets; red circles show results for fits that do not include the measured $\sigma_{DC}$.

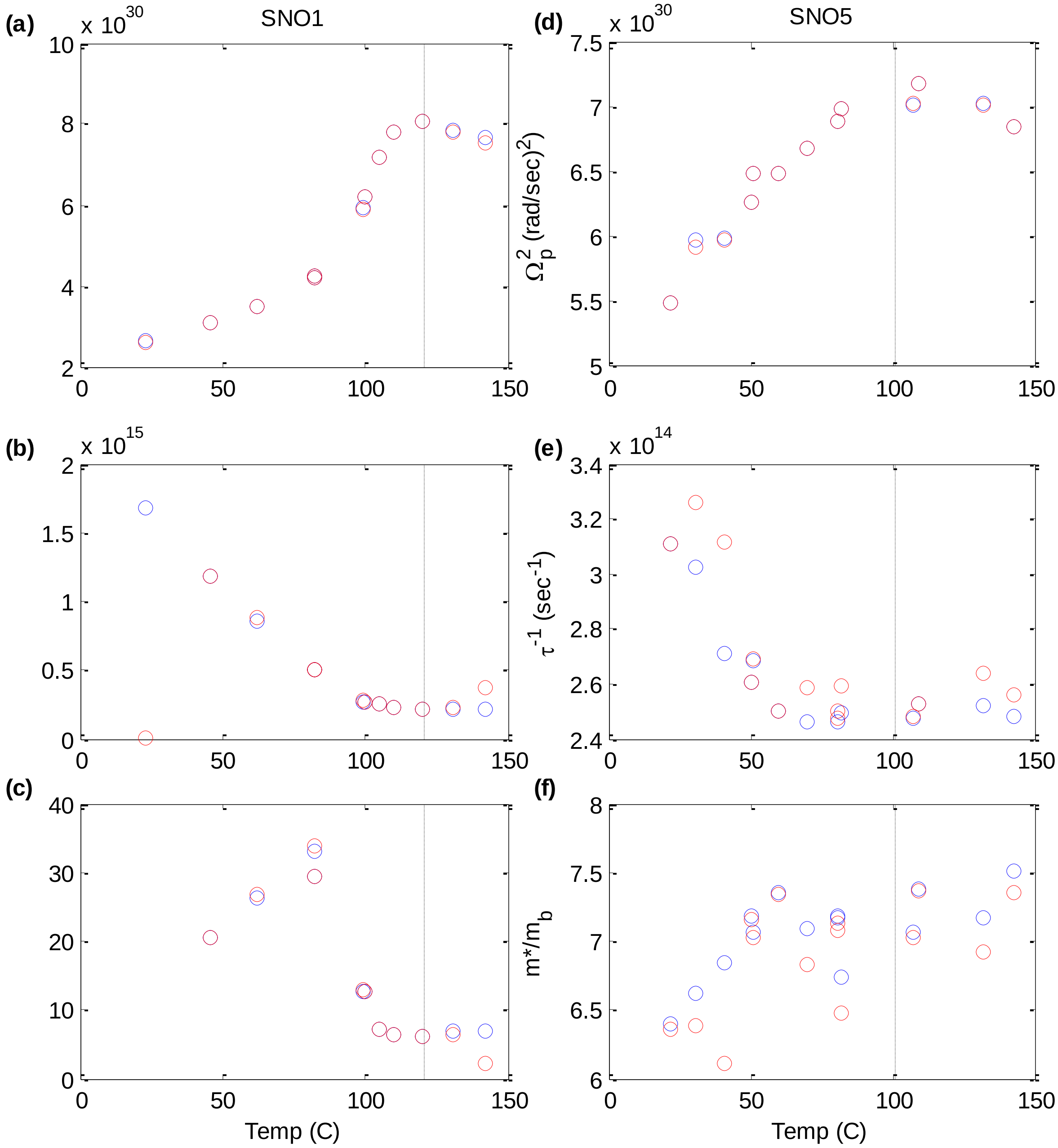

**Figure S5:** Comparing FTIR fits with and without including the measured $\sigma_{DC}$ in the modeled datasets. (a-c) SNO5. (d-f) SNO1. Blue circles show results for FTIR fits that include the measured $\sigma_{DC}$ in the modeled datasets; red circles show results for fits that do not include the measured $\sigma_{DC}$. Dashed vertical line marks $T_{IM}$. Top row: $\Omega_p^2$ evaluated at 3000 cm$^{-1}$. Middle row: scattering rate $\tau^{-1}$ in the $\omega \rightarrow 0$ limit from the extended Drude analysis. Bottom row: $m^*/m_b$ in the $\omega \rightarrow 0$ limit from the extended Drude analysis.

## 3. Estimating the total carrier density ($n$), $\sigma_{MIR}$, and $m^*$

$\sigma_{MIR}$ is usually defined by the condition that the scattering length is equal to the distance $d$(Ni-Ni) between cations. This can be written as

$$\rho = \frac{3\pi^2\hbar}{q^2 k_F^2 l} \tag{S1}$$

where $k_F$ is the Fermi wavevector and we set the scattering length $l = d$ [1]. For a negative charge transfer material such as $R$NiO$_3$ it might be more appropriate to instead use the distance $d$(O-O) between anions, but this is nearly identical to $d$(Ni-Ni). We use the pseudocubic lattice constant $d$ = 3.8 Å. In order to calculate $k_F$ we need to estimate the total metallic carrier density $n$. In the PM phase $R$NiO$_3$ is partially compensated, with both electron-like and hole-like states at $E_F$, and therefore we cannot simply interpret our Hall effect measurements in terms of total carrier density[17]. Instead we estimate $n$ by assuming 1 carrier per Ni, giving $n = 1.8\text{x}10^{22}$ cm$^{-3}$. This is a common assumption in the field and is consistent with what is known about the low-energy electronic structure (*c.f.* Fig. 4)[27]. With these estimates and the expression $k_F = \left(3\pi^3 n\right)^{1/3}$ we calculate $\sigma_{MIR} \approx 2000\ \Omega^{-1}$ cm$^{-1}$. This is slightly higher than the frequently quoted estimate $\sigma_{MIR} \approx 1000\ \Omega^{-1}$ cm$^{-1}$ [2], which is appropriate for materials with higher carrier density.

We can combine the above estimate of $n$ with our FTIR results to estimate the effective mass $m^*$ (see Eq. 2). For all samples we find $m^* \approx 7m_0$ for $T > T_{IM}$, in agreement with previous estimates and typical of a moderately correlated metal[5,14].

## 4. Resistivity at low temperature

The insulating energy gap ($E_g$) and the conduction mechanism below $T_{IM}$ have long remained unknown, due in large part to the failure of any single mechanism to model $\sigma_{DC}(T<T_{IM})$. To demonstrate this we plot in Fig. S6 $\rho(T)$ and $S \equiv -d\log_{10}\rho / d\log_{10}T$ measured on SNO3 for $T$ > 2 K. Data are measured using sputtered Pt electrodes. For common conduction mechanisms $S(T)$ has a fixed slope ($m$) on a log-log plot: $m$=1/4 for variable range hopping (VRH, also called Mott variable range hopping), $m$=1/2 for VRH in the presence of Coulomb interactions (also called Efros-Shklovskii VRH), and $m$=1 for activated conduction. Instead the data have a $T$-dependent slope that is inconsistent with a single mechanism over any appreciable range. Of these three mechanisms the data most closely approximate Efros-Shklovskii VRH, but only in the limited range of 4 – 10 K. The data are also incompatible with activated conduction with a $T$-dependent $E_g$ with the critical form $E_g = E_{g0}\left(T_{IM} - T\right)^{\beta}$.

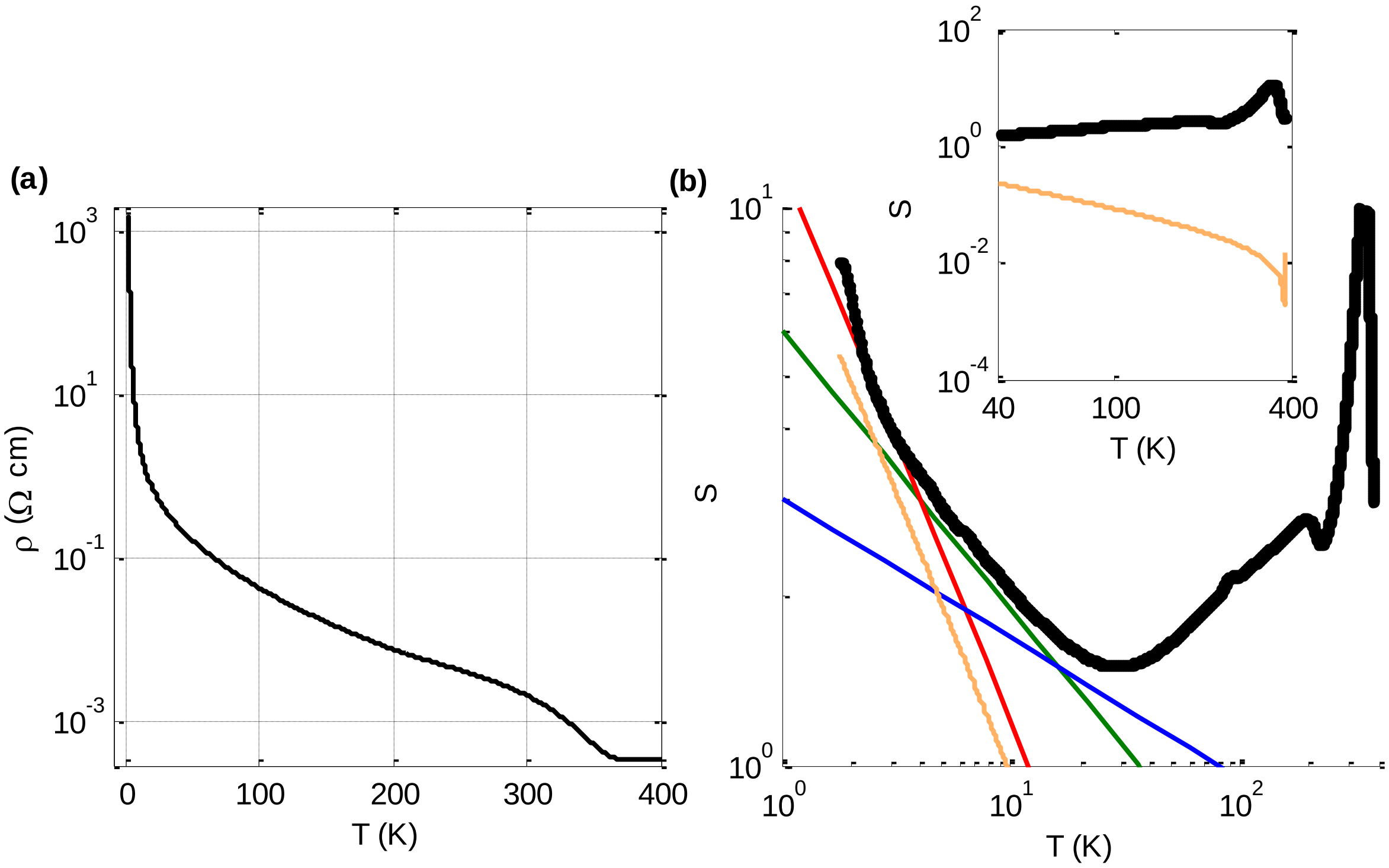

**Figure S6:** ρ($T$) at low $T$ and the conduction mechanism in the insulating phases. (a) ρ($T$) for SNO3 measured for 2 – 400 K. (b) $S \equiv -d\log_{10}\rho / d\log_{10}T$ (black points) calculated from the data in (a). The colored lines show the slopes ($m$) expected for variable range hopping ($m$ = 1/4, blue line), variable range hopping in the presence of Coulomb interactions ($m$ = 1/2, green line), and activated conduction ($m$ = 1, red line). The orange line shows the expected slope for activated conduction with a $T$ dependent bandgap ($E_g$): $E_g(T) = E_{g0}(T_{IM} - T)^{\beta}$ with $T_{IM}$ = 380 K. (Inset) Data and $E_g(T)$ model from (a), with range expanded to show difference in slopes between the data and the model for $T \to T_{IM}^{-}$

.

## 5. References for Fig. 1: published conductivity in the high $T$ PM phase of the nickelates

$LaNiO_3$ (+): Rajeev *et al.*[32]

$LaNiO_{3-\delta}$, δ = 0.02 (○): Gayathri *et al.*[33]

$LaNiO_{3-\delta}$, δ = 0.14 (□): *ibid.*

$LaNiO_3$ (☆): Pal *et al.*[34]

$La_{1-x}Pb_xNiO_3$, x = 0.1 (✡): *ibid.*

$PrNiO_3$ (×): Novojilov *et al.*[35]

$NdNiO_3$ (◁): Liu *et al.*[36]

$NdNiO_{3-\delta}$, $\delta = 0.06$ (■): Nikulin *et al.*[37]

$NdNiO_{3-\delta}$, $\delta = 0.29$ (◆): *ibid*.

$NdNiO_3$ (▲): Cheong *et al.*[38]

$NdNiO_3$ (▷): Vassiliou *et al.*[39]

$Sm_xNd_{1-x}NiO_3$, x = 0.2 (◇): Capon *et al.*[40]

$NdNiO_3$ (▽): Ambrosini and Hamet[41]

$Sm_xNd_{1-x}NiO_3$, x = 0.4 (△): *ibid.*

$SmNi_{1-x}Co_xO_3$, x = 0.03 (▶): Perez-Cacho *et al.*[42]

$SmNi_{1-x}Co_xO_3$, x = 0.075 (◀): *ibid.*

$SmNiO_3$ (▼): *ibid.*

$SmNiO_3$ (✳): Conchon *et al.*[43]

$GdNiO_3$ (●): Novojilov *et al.*[44]

# Methods

## 1. $SmNiO_3$ thin film growth and characterization

We grow $SmNiO_3$ (SNO) thin films on $LaAlO_3$ (LAO) substrates by reactive sputtering from $SmNiO_x$ ceramic target (ACI Alloys). Our substrates are LAO single crystals (MTI Corporation), <001> pseudocubic orientation, 10x10x0.5 $mm^3$, with a polished front side and a rough back side. In order to obtain the correct phase it is necessary to sputter in a high background oxygen pressure[16]. The films SNO1-SNO5 studied here were grown with substrate temperature $T_{sub}$ = 650 °C and background pressure in the range 260 – 380 mTorr (highest for SNO1, lowest for SNO5) with a gas mixture of 80/20 sccm Ar/$O_2$. By varying the background pressure we control the oxygen stoichiometry and the electronic properties[16,17]. Most *R*NiO3 samples, both here and in the literature, are in fact $RNiO_{3-\delta}$ ($\delta > 0$) due to the well-known difficulties in synthesizing fully stoichiometric $RNiO_3$ [45]. Oxygen-rich films such as SNO1 exhibit sharp insulator-metal transitions. As $\delta$ increases the IMT becomes less distinct. Oxygen-poor films such as SNO5 have higher resistivity in the metal phase and lower resistivity in the insulating phases. The same trends have been observed with Co doping in $SmNi_{1-x}Co_xO_3$, Ca doping in $Sm_{1-x}Ca_xNiO_3$, and oxygen reduction in $NdNiO_{3-\delta}$ [17]. Therefore oxygen vacancies are shallow acceptors in the insulating phase and scattering sites in the metallic phase.

Due to twinning of the LAO substrates we cannot measure unambiguous reciprocal space maps. However, our x-ray data are consistent with fully strained epitaxial films. SNO is compressively strained on LAO with a -0.15% (pseudocubic) lattice mismatch. Compressive strain suppresses $T_{IM}$ in the nickelates. For bulk SNO $T_{IM}$ = 400 K. For our films $T_{IM}$ varies from 374 to 394 K (SNO5 and SNO1, respectively), consistent with published reports of epitaxial, strained SNO on LAO[46].

We measure film thickness using x-ray reflectivity. Our films are 15.7, 15.5, 18.8, 16, and 19.4 nm thick (SNO1, SNO2,… SNO5). Atomic force microscopy (AFM) shows that our films are atomically smooth and continuous. In Fig. S7 we show AFM height images for SNO1 and SNO5. In both images the film roughness is below 200 pm and the atomic terraces of the underlying LAO substrates are plainly visible.

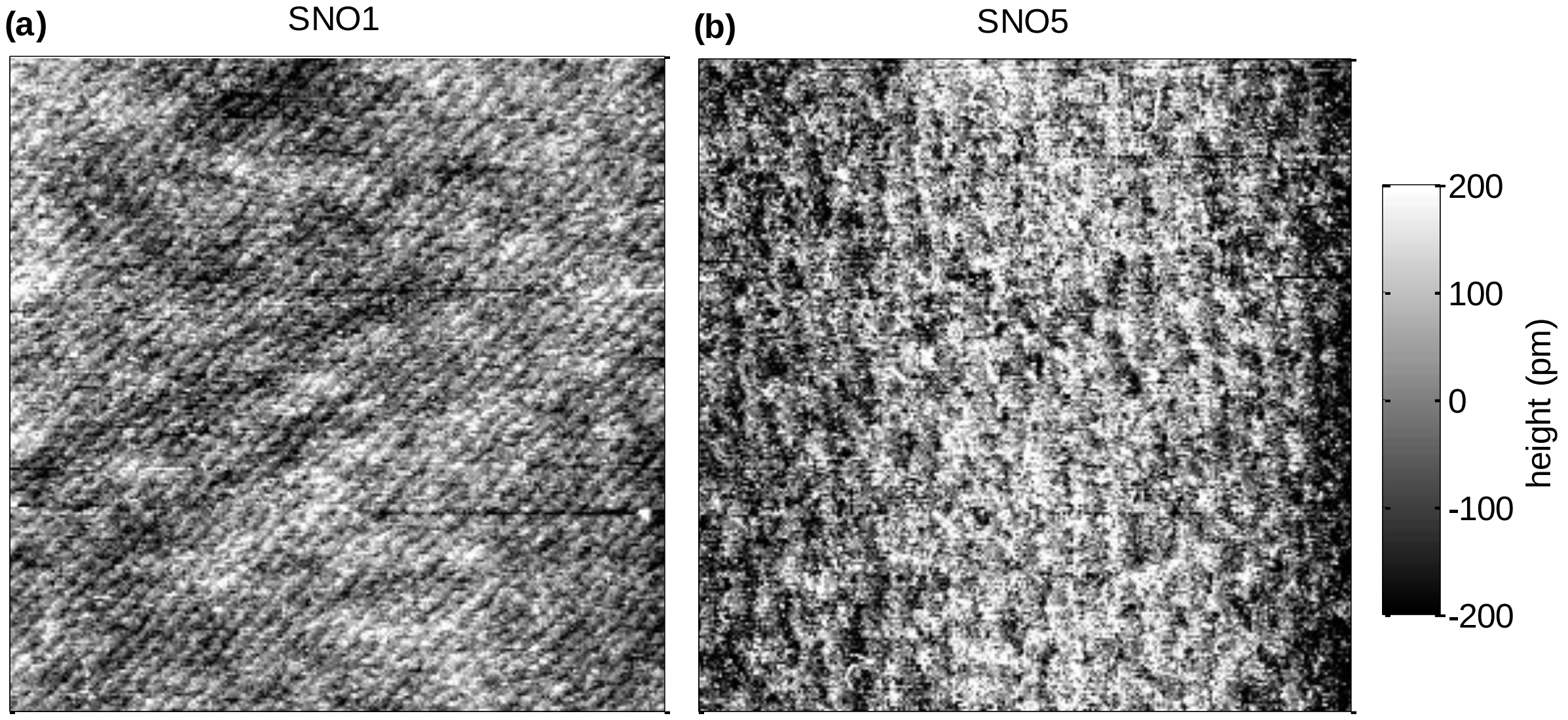

**Fig. S7:** AFM height images for (a) SNO1 and (b) SNO5. Both scans are 5x5 $\mu m^2$ and the height color scale is the same. The ridges are atomic step edges from the underlying LAO substrates. The RMS roughness of both films is approximately 120 pm.

## 2. Fourier transform infrared (FTIR) reflectivity measurements

FTIR data was recorded using a ThermoFisher Nicolet 6700 spectrometer. We measured near-normal reflectivity (*R*) with fixed incidence angle of 10° using a Harrick Scientific reflectivity stage. Sample temperature (*T*) was varied using a home-built thermal stage that was mounted on top of the reflectivity stage. All samples were mounted to a copper block using thermal conductive paste, and *T* was measured by a thermocouple mounted on the same block several mm away from the sample. Data was recorded on both warming and cooling for all samples, and we observe no thermal hysteresis. For all *R* data we use a gold mirror as the *R* = 100% standard. We used substrates with an unpolished back side in order to suppress the effects of rear-surface reflection in the IR frequency windows in which LAO is transparent.

Data is recorded in three separate energy ranges using a selection of sources, beam splitters, and detectors: far-IR (50-600 $cm^{-1}$), mid-IR (400-4000 $cm^{-1}$), and near-IR (1200-12500 $cm^{-1}$). At the far end of the far-IR spectrum the signal is very weak, and the data is noise limited and non-reproducible. Therefore we restrict our analysis to the range 200-12500 $cm^{-1}$ over which our data is reproducible. We analyze all raw data including the overlapping ranges (400-600 $cm^{-1}$ and 1200-4000 $cm^{-1}$) without scaling, after dark correction (see below).

For heated samples FTIR data should to be corrected for the effects of sample emission when the sample and the detector are at different *T* [47–49]. We correct for sample emission by recording a spectrum with the IR source switched off and allowed 10 min to cool. This "dark" spectrum is then subtracted from the spectrum recorded with the IR source on. All of our data for $T > 50$ °C is corrected in this way. The fractional correction ($R$(lamp off)/$R$(lamp on)) only becomes experimentally significant for $T > 100$ °C. For all $T < 100$ °C, $R$(lamp off)/$R$(lamp on) is below 0.02. The required source cooling time for acquiring dark spectra was independently measured by heating a gold reference mirror and fitting the measured dark spectra by Planck model for black body emission.

## 3. Temperature-dependent optical model for the $LaAlO_3$ substrates

To accurately model the optical spectrum of our films we need a model of the substrate dielectric function $\varepsilon(\omega)$. The IR dielectric response of LAO at room $T$ is reported in several publications, but there is no available detailed report of the $T$-dependence in our experimental range. Therefore we constructed a $T$ dependent LAO dielectric function model starting from published results at room $T$ and extended by $T$ dependent measurements of our bare substrates.

We measure $R$ over the full energy and $T$ range of our experiment for bare LAO substrates, including dark correction for $T > 50$ °C. We make repeated measurements using 2 different (nominally identical) bare substrates, for a total of 12 $T$ points between 22 and 142 °C. We measure the low frequency dielectric constant ($\varepsilon(\omega \rightarrow 0)$) over the full $T$ range of our experiment using capacitor test structures fabricated from bare LAO substrates. We find that $\varepsilon(\omega \rightarrow 0) = 26.4 \pm 0.5$ and is independent of frequency and $T$ in the ranges 1 kHz – 2 MHz and 22 – 142 °C.

To model our data we start with the dielectric function for LAO at room $T$ reported by Zhang[50]. We model the 3 major phonon modes using the factorized TO-LO expression for $\varepsilon(\omega)$, and we model the weaker phonons as Lorentz oscillators (note that Zhang models all modes as Lorentz oscillators)[51]. To the model of Zhang we add two additional weak phonon modes near 593 and 666 $cm^{-1}$ that are visible in our data. We fit our data using this model and compile the resulting fit parameters for all measurements. All fits are constrained such that $\Omega_{LO} > \Omega_{TO}$ and $\gamma_{TO}/\gamma_{LO} \geq (\Omega_{TO}/\Omega_{LO})^2$, as required for the factorized TO-LO dielectric function model [51]. From the compiled fit parameters we construct a $T$ dependent model for $\varepsilon(\omega)$ of LAO.

In Fig. S8 we plot representative data and the fit. Our $T$ dependent model is presented in Table S1. The error bars in the parameters are derived from the scatter in the fit parameters from multiple datasets.

Our films are grown with substrate temperature $T_{sub} = 650$ °C and background pressure in the range 260 – 380 mTorr with a gas mixture of 80/20 sccm $Ar/O_2$. We tested the effects of this environment on the LAO dielectric function by exposing a bare substrate to a typical growth run, but with the growth shutter closed, and then measuring $R$ at room $T$. We observed no change.

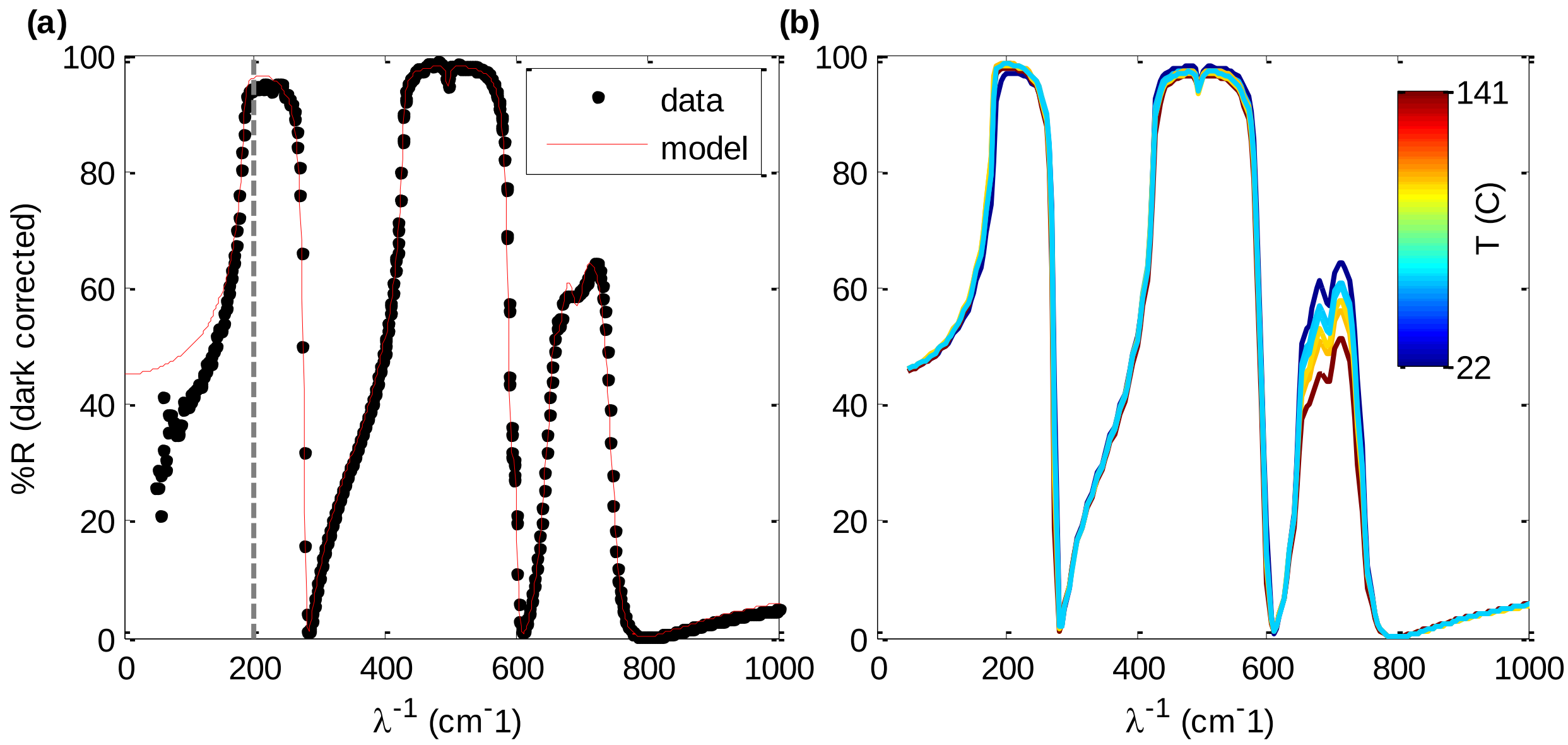

**Fig. S8:** *T* dependent dielectric constant of LAO. (a) *R* data and fit for LAO at 22 °C. We only show data up to 1000 $cm^{-1}$ to emphasize the phonon spectrum; at higher energy there are no phonons and *R* is largely controlled by $\varepsilon_\infty$. We only fit data for $\lambda^{-1} > 200$ $cm^{-1}$ (vertical dashed grey line) due to weak signal and non-reproducible data at the lowest energies. (b) *R* at select *T* calculated for LAO using our dielectric model.

| Phonon | Parameter | Value |
|---|---|---|
| n/a | $\varepsilon_\infty$ | 4.51 ± 0.1 |
| LO-TO #1 | $\Omega_{TO}$ | 187.3 ± 2.3 |
| | $\Omega_{LO}$ | 278.8 ± 1.3 |
| | $\gamma_{TO}$ | 6.7 ± 2.7 |
| | $\gamma_{LO}$ | 6.9 ± 2.3 |
| LO-TO #2 | $\Omega_{TO}$ | 427.7 ± 0.6 |
| | $\Omega_{LO}$ | 607 ± 6.2 |
| | $\gamma_{TO}$ | 3.7 + 0.0173**T* ± 2 |
| | $\gamma_{LO}$ | 3.74 + 0.0185**T* ± 2 |
| LO-TO #3 | $\Omega_{TO}$ | 650.8 ± 0.5 |
| | $\Omega_{LO}$ | 730.9 – 0.0836**T* ± 2 |
| | $\gamma_{TO}$ | 13.71 + 0.055**T* ± 1 |
| | $\gamma_{LO}$ | 13.7 + 0.0563**T* ± 1 |
| LOR #1 | $\Omega_0$ | 495.9 – 0.013**T* ± 0.2 |
| | $\Omega_p$ | 37.4 ± 4.7 |
| | $\gamma$ | 4.3 ± 1.5 |
| LOR #2 | $\Omega_0$ | 594.2 ± 0.9 |
| | $\Omega_p$ | 40.4 ± 7.3 |
| | $\gamma$ | 10.8 ± 1.6 |
| LOR #3 | $\Omega_0$ | 666.4 + 0.0294*$T$ ± 1.5 |
| | $\Omega_p$ | 189.1 ± 15.7 |
| | $\gamma$ | 21.6 + 0.0516*$T$ ± 1.5 |
| LOR #4 | $\Omega_0$ | 694.5 + 0.012*$T$ ± 1 |
| | $\Omega_p$ | 94.2 + 0.2*$T$ ± 4 |
| | $\gamma$ | 22.7 + 0.0229*$T$ ± 1 |

**Table S1:** LAO dielectric function model for 22 < T < 142 °C. All model parameters are in units of $cm^{-1}$ except for $\varepsilon_\infty$, which is unitless; $T$ is in units of °C. Model consists of 3 factorized LO-TO oscillators (LO-TO #1-3) and 4 Lorentz oscillators (LOR #1-4). For the LO-TO models: $\Omega_{TO}$, $\Omega_{LO}$ = transverse, longitudinal phonon frequencies; $\gamma_{TO}$, $\gamma_{LO}$ = transverse, longitudinal phonon damping. Fits must be constrained so that $\Omega_{LO} > \Omega_{TO}$ and $\gamma_{TO}/\gamma_{LO} \geq (\Omega_{TO}/\Omega_{LO})^2$. For the LOR models: $\Omega_0$ = phonon frequency; $\Omega_p$ = plasma frequency; $\gamma$ = damping.

## 4. FTIR Data Analysis for $SmNiO_3$ thin films on $LaAlO_3$ substrates

We use Drude-Lorentz modeling to fit the $R$ data and extract the dielectric function $\varepsilon(\omega)$ of our thin film samples. Analysis is done using the program RefFIT (Alexey Kuzmekno, University of Geneva; http://optics.unige.ch/alexey/reffit.html). We model our sample structure taking into account fully coherent reflections at the film-substrate interface and fully incoherent reflections at the substrate-air interface. The film thickness is independently measured by x-ray reflectivity and is a model input. The optical response of the LAO substrate is taken from the model described in Section 3 (above). The LAO model parameters are allowed to vary within the bounds shown in Table S1 while being constrained so that $\Omega_{LO} > \Omega_{TO}$ and $\gamma_{TO}/\gamma_{LO} \geq (\Omega_{TO}/\Omega_{LO})^2$.

For modeling the film dielectric function we start with published results for SNO at 300 K reported by Mroginski[11]. Ref. [11] uses LO-TO modeling, which we convert to Lorentz modeling with a single damping parameter. Starting with this model we add or remove oscillators as needed to produce a stable and satisfactory fit. Ref. [11] reports measurements on bulk SNO ceramics, and achieves better resolution of the phonon spectrum than we are able. Our best fits therefore contain fewer phonon modes than reported in Ref. [11]. The most prominent features in our raw data result from the substrate, and are modulated by the conductivity of the overlying films. However the strong SNO bending- and breathing-mode phonons that are important for our analysis are usually visible in the raw data and are easily modeled. In addition to phonon modes we need one broad oscillator each in the mid- and near-IR in order to model our data. We use a single Drude term for all fits. For all fits $\varepsilon_\infty$ is fixed at 2. This is in agreement with reported $T$ dependent values for $\varepsilon_\infty$ of SNO, and constraining $\varepsilon_\infty$ improves greatly the stability and reproducibility of the fits[11].

For all fits presented (and unless otherwise noted) we have included the independently measured $\sigma_{DC}(T)$ in the modeled dataset. $\sigma_{DC}(T)$ was measured on the same films on a probe station using sputtered Pt contacts in the van der Pauw geometry. $\sigma_{DC}$ is weighted to account for possible differences in $T$ calibration between different temperature stages, an effect which is significant near $T_{IM}$ where $\sigma_{DC}(T)$ varies rapidly. Importantly, the decision to include or not to include $\sigma_{DC}(T)$ in the fits causes only minor quantitative changes in $\sigma'(\omega)$, $f(\hbar\omega)$, extended Drude analysis and $\Omega_p^2$. Our results are not affected by the choice of including $\sigma_{DC}(T)$ in the fits. This is described more fully in the Supplementary Discussion.

In Fig. S9 we show the data and the fits for representative datasets for all samples. The data are somewhat messy near 2000 cm$^{-1}$ due to the combined contributions of the backside reflection and the $H_2O$ absorption lines which sometimes appear due to incomplete purging of the FTIR bench. We prevent these features from affecting the fits by sampling more coarsely in this range, and by preventing the fitting routine from placing any Lorentz oscillators corresponding to these narrow features.

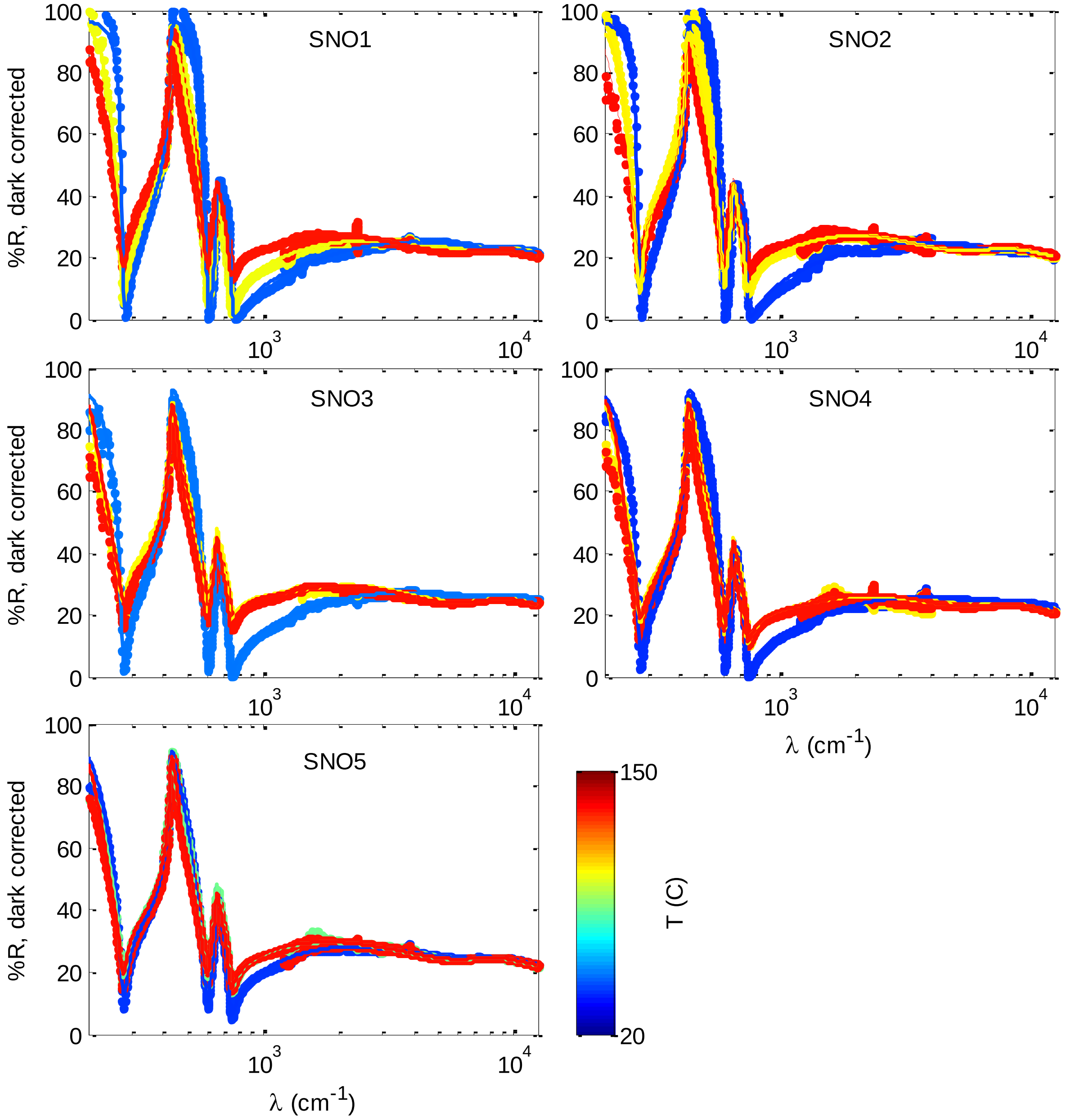

**Fig. S9:** *R* data and fits for select *T* for all samples. For each sample data (points) and fits (lines) are plotted for 3 temperatures near 40, 100, and 130 °C. The colorbar indicates *T* for all plots.